# Imaging a 1-electron InAs quantum dot in an InAs/InP nanowire

*A. C. Bleszynski[1], L.E. Fröberg[2], M. T. Björk[2], H. J. Trodahl[1]*

*L. Samuelson[2] and R.M. Westervelt[1]*

[1]Department of Physics and Division of Engineering and Applied Sciences,

Harvard Univ., Cambridge, Massachusetts 02138, USA.

[2]Solid State Physics/the Nanometer Structure Consortium,

Lund Univ., Box 118, S-221 00 Lund, Sweden.

Nanowire heterostructures define high-quality few-electron quantum dots for nanoelectronics, spintronics and quantum information processing. We use a cooled scanning probe microscope (SPM) to image and control an InAs quantum dot in an InAs/InP nanowire, using the tip as a movable gate. Images of dot conductance *vs.* tip position at T = 4.2 K show concentric rings as electrons are added, starting with the first electron. The SPM can locate a dot along a nanowire and individually tune its charge, abilities that will be very useful for the control of coupled nanowire dots.

Semiconducting nanowire heterostructures [*1*] provide an excellent system in which to make high-quality ultra-small quantum dots for applications in nanoelectronics, spintronics and quantum information processing (QIP) [*2*,*3*]. The bottom up nature of the construction of quantum dots in nanowire heterostructures [*1*,*4*,*5*] results in atomically sharp interfaces and highly controllable dot size, shape, and composition. Few-electron nanowire quantum dots have been recently reported [*6*,*7*] that exhibit a well-defined atom-like electronic shell structure down to the last electron [*8*,*9*]. The ability to operate nanowire quantum dots in the one-electron regime makes them attractive candidates for QIP.

InAs is a particularly attractive material for several reasons. InAs has a large g-factor, making it useful for spintronic and QIP devices [*2*,*3*]. Furthermore, the g-factor of an InAs nanowire quantum dot can be varied from 2 to the bulk value 14, by varying the dot size [*10*], a consequence of quantum confinement [*11*,*12*]. Lastly, whereas some semiconductors have surface depletion, the surface of InAs is known to have a charge accumulation layer [*13*]. This potentially allows for very small diameter nanowires and ultrasmall dots that are not depleted of electrons, as well as for Schottky-barrier-free contacts to metallic leads.

In order to control the charge in an individual nanowire quantum dot in a dot circuit, new gating techniques will be needed. A conventional back gate couples to the entire nanowire and all of the dots in it. A lithographically defined gate for an individual dot has to be small and aligned with high precision, which is particularly difficult for heterostructure nanowire quantum dots, because of their small size and the uncertainty in their location. A scanned probe microscope (SPM) can overcome both of these obstacles by using the conducting tip as a movable gate. Cooled SPMs have proven to be powerful tools for imaging the electronic properties of nanoscale systems including nanotubes, nanowires and two-dimensional electron gas structures [*14-19*], and they can image the presence of a single electronic charge [*17-18,20-23]*. In previous work, a cooled SPM was used to image quantum dots unintentionally formed in carbon nanotubes [*17,18*] and in semiconducting nanowires [24].

In this letter, we present scanning gate images at 4.2 K of a one-electron InAs quantum dot formed in an InAs/InP nanowire heterostructure by two InP barriers. We use the conducting tip of a cooled SPM as movable gate to locate the quantum dot and tune its charge, starting with the first electron. The images show concentric rings of high conductance about the dot corresponding to Coulomb blockade peaks as an electron is added. In this way, the SPM can locate a dot along the nanowire and individually tune its charge. This ability will be very useful for the manipulation of multiple coupled quantum dots grown along a nanowire heterostructure.

Figure 1(a) shows the SPM imaging setup. Using a Coulomb-blockade imaging technique [*17-18,20-23]*, the charged SPM tip is scanned in a plane above the nanowire, and the resulting change in nanowire conductance $G$ is recorded to form a two-dimensional image at 4.2 K. Modeling the dot as a small metal sphere, the charge induced by the tip is $q_{dot}(V_t, r_t) = C_{t-d}(r_t) * V_{t-d}$, where $C_{t-d}(r_t)$ is the tip-dot capacitance at a distance $r_t$ away and $V_{t-d} = V_{tip} + V_{cont}$ is the voltage difference between the tip and the dot for tip voltage $V_{tip}$ including the contact potential $V_{cont}$. We assume $C_{t-d}(r_t) << C_\Sigma$ where $C_\Sigma$ is total dot capacitance to ground. Scanning the tip with fixed $V_t$ changes $C_{t-d}(r_t)$, which varies $q_{dot}(V_t, r_t)$ and causes oscillations in dot conductance $G$ each time an electron is added to the dot. In the images, this behavior manifests itself as concentric rings of peaked conductance surrounding the quantum dot with each ring corresponding to a Coulomb blockade peak. The rings thus locate the quantum dot. If the tip is scanned along one of these rings, the induced charge $q_{dot}(V_t, r_t)$ remains constant - the rings are contours of constant tip-dot coupling.

The nanowires used in this experiment were catalytically grown from Au nanoparticles on an InAs <111> B substrate using chemical beam epitaxy [*25*]. A TEM image of a typical nanowire quantum dot is shown in Fig. 1(b): the dark sections are InAs and the light sections are InP. The InAs/InP heterostructure is formed by alternating the gas precursors during the growth process. The diameter of the wires is ~ 50 nm and their lengths are ~ 3 μm. An 18 nm long InAs quantum dot is formed between two 8 nm thick InP barriers. The 600 meV conduction band offset between InAs and InP produces electron confinement inside the InAs quantum dot. After growth, the nanowires are deposited onto a degenerately doped Si substrate capped with a 100 nm $SiO_2$ layer. This conducting substrate acts as a non-local back gate that, through an applied voltage $V_{bg}$, can tune the Fermi level in the nanowire. Ni/Au electrode contacts, spaced by 2 μm, are defined with e-beam lithography as indicated in Fig. 1(a). The thickness of the InP barriers and the InAs dot are tuned such that the few-electron Coulomb blockade regime can be reached for small $V_{bg}$.

The number of electrons on the dot and the energy of the first few electron states can be determined from the Coulomb blockade diamonds, plots of $G$ vs. gate voltage and source-to-drain voltage $V_{sd}$ shown in Fig. 2. Figure 2(a) was recorded by sweeping $V_{sd}$ and the backgate voltage $V_{bg}$ while the tip position was fixed 50 nm above the dot with constant tip voltage $V_{tip} = -2.0V$ ; the dot is emptied of electrons for $V_{bg} < 0.4V$ due to quantum confinement in the growth direction [*8*]. The plot displays regions of zero conductance, Coulomb diamonds, when $V_{sd}$ is smaller than the energy required to add another electron to the dot. The diamond size is seen to vary with electron number, revealing a shell structure of electronic states in the quantum dot [*9*].

A similar set of Coulomb diamonds shown in Fig. 2(b) was obtained by fixing the tip position 70 nm directly over the dot and sweeping $V_{tip}$ and $V_{sd}$ with fixed $V_{bg} = 1.2V$ . The dot is emptied of electrons for $V_{tip} < -2.0V$ . From the change in tip voltage required to add one electron we find $C_{t-d}(r_t) \sim 0.4aF$ for tip distance $r_t \sim 70nm$. Like the back gate, the tip can tune the dot's charge to zero electrons. Unlike the back gate, the tip offers the extra advantage of movability: the tip's coupling to the dot can be varied through positioning. The sizes of the Coulomb diamonds change with different combinations of tip and backgate voltages $V_{tip}$ and $V_{bg}$, as shown in Figs. 2(a) and 2(b). This occurs because $V_{tip}$ and $V_{bg}$ induce charge in the nanowire dot by shifting the electron density profile sideways, toward or away from the substrate; this compresses the wavefunction against the sides of the nanowire and changes the energy of quantum states. For different fixed backgate voltages, we obtained differently sized Coulomb diamonds in tip-voltage plots similar to Fig. 2(b).

SPM images of the last electron on the quantum dot are shown in Figs. 3(a-b). We adjust $V_{bg}$ and $V_{tip}$ so the dot is tuned to the zero-one electron transition when the tip is nearby. The images of *G vs.* tip position in Figs. 3(a-b), recorded as the tip is scanned in a plane above the nanowire with fixed tip voltage $V_{tip}$, display a ring centered on the InAs dot that corresponds to the Coulomb-blockade conductance peak as the first electron is added the dot. When the tip is inside the ring in Fig. 3(a), the dot is empty, and when the tip is just outside the ring, the dot holds one electron. The Coulomb blockade ring for the addition of the second electron is visible at the corners of Fig. 3(a). As $V_{tip}$ is made less negative in Fig. 3(b), the first Coulomb blockade ring shrinks to a point and the ring for the second electron also shrinks in size.

Figures 3(c-d) show SPM Coulomb blockade conductance images of the InAs dot when it is tuned to hold a larger number of electrons. Again, rings of peaked conductance surround the quantum dot. Moving radially outwards, each new ring corresponds to the addition of an electron to the dot, as indicated by the integers. As $V_{bg}$ is changed to a more positive value in Fig. 3(d), the rings shrink in radius, because the back gate pulls more electrons onto the dot. To obtain the same number of electrons, the induced charge $q_{dot}\left(V_{tip},r_t\right)$ must be made more negative by moving the tip closer.

For certain combinations of $V_{tip}$ and $V_{bg}$, the SPM images, such as Fig. 4, show an additional set of narrowly spaced Coulomb conductance rings centered about a section of the nanowire to the left of the grown-in InAs quantum dot, clear evidence of the formation of a second quantum dot. By comparing the spacing of the rings surrounding the two dots, the length of the extra dot is found to be ~ 200 nm, [24] comparable to the length between the grown-in dot and the contact as shown.

One-electron double dots, grown inside an InAs/InP nanowire, are attractive for spin manipulation, because the dots can be very small and closely spaced. However, it is difficult to gate each dot individually, because the dot size and spacing are often smaller than the spatial resolution of e-beam lithography. Fuhrer *et al.* [26] used an array of many gates to characterize nanowire double dots, by carefully measuring the dot-gate couplings and tuning the gate voltages accordingly.

Using a conducting SPM tip, we should be able to individually tune the charge in of each dot in an InAs double dot grown in an InAs/InP nanowire. Figures 5(a-b) show simulated SPM conductance images of two InAs quantum dots, each with length 25 nm and diameter 50 nm, defined by 5 nm InP barriers. The nanowire lies on its side on a substrate, and an SPM tip is scanned in a plane 25 nm above its diameter with constant tip voltage. The number of electrons on a given dot increases in integer steps with tip-dot radius to form a bullseye-shaped diagram, as described above; rings of high conductance occur at radii where the electron number changes. For a double dot, the centers of the two bullseyes are at different tip positions, as shown in Fig. 5(b). The SPM conductance image locates the double dot along the nanowire, and by simply moving the tip, one can individually tune the charge on each dot. Conductance through two dots in series occurs when both dots conduct, so conductance in the image of Figs. 5(a-b) only occurs at the intersections of the conductance rings for each dot.

Conductance images of nanowire double dots *vs.* like Fig. 5 are equivalent to traditional two-dimensional plots of double-dot conductance *vs.* gate voltage for two lithographically defined gates. Using an SPM tip as a gate should allow a full range of experiments to manipulate charges and spins on nanowire double dots, that take advantage of the small dot size, large g-factor, and relatively high operating temperature.

We thank M. Stopa and E. J. Heller for helpful suggestions. This research was supported at Harvard by the NSF-funded Nanoscale Science and Engineering Center under grant NSF-PHY-0117795 and the Army Research Office under grant ARO-W911NF-04-1-0343, and at Lund University by the Swedish Research Council (VR), the Swedish Foundation for Strategic Research (SSF), the Knut and Alice Wallenberg Foundation (KAW), the Office of Naval Research (ONR) and by the EU program NODE 015783.

FIGURES

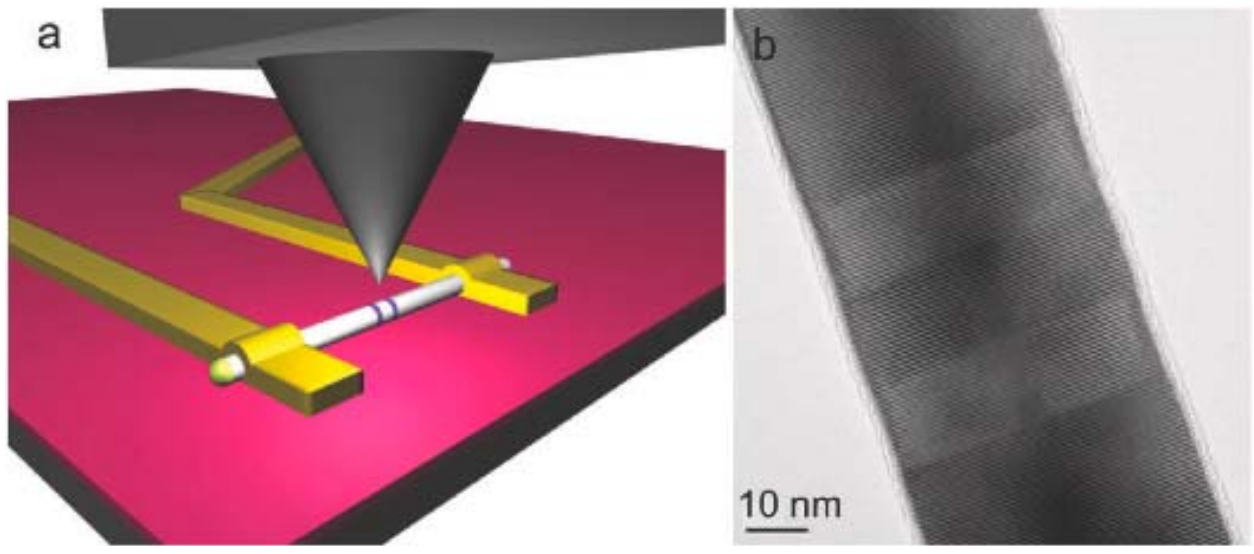

FIG. 1 (a) Experimental setup. A metallized scanning probe microscope (SPM) tip is scanned at a fixed height above a nanowire. Conductance between source and drain electrodes is recorded *vs.* tip position to obtain an image. The tip is scanned in a plane typically 20 nm to 100 nm above the nanowire with tip voltages -3V to 3V. (b) TEM image of an InAs/InP heterostructure nanowire similar to the ones used in this experiment. Individual atomic layers are clearly visible, indicating the high quality of the epitaxial growth. An InAs quantum dot (dark) with a well defined disc geometry is confined between two InP barriers (light) with InAs leads (dark).

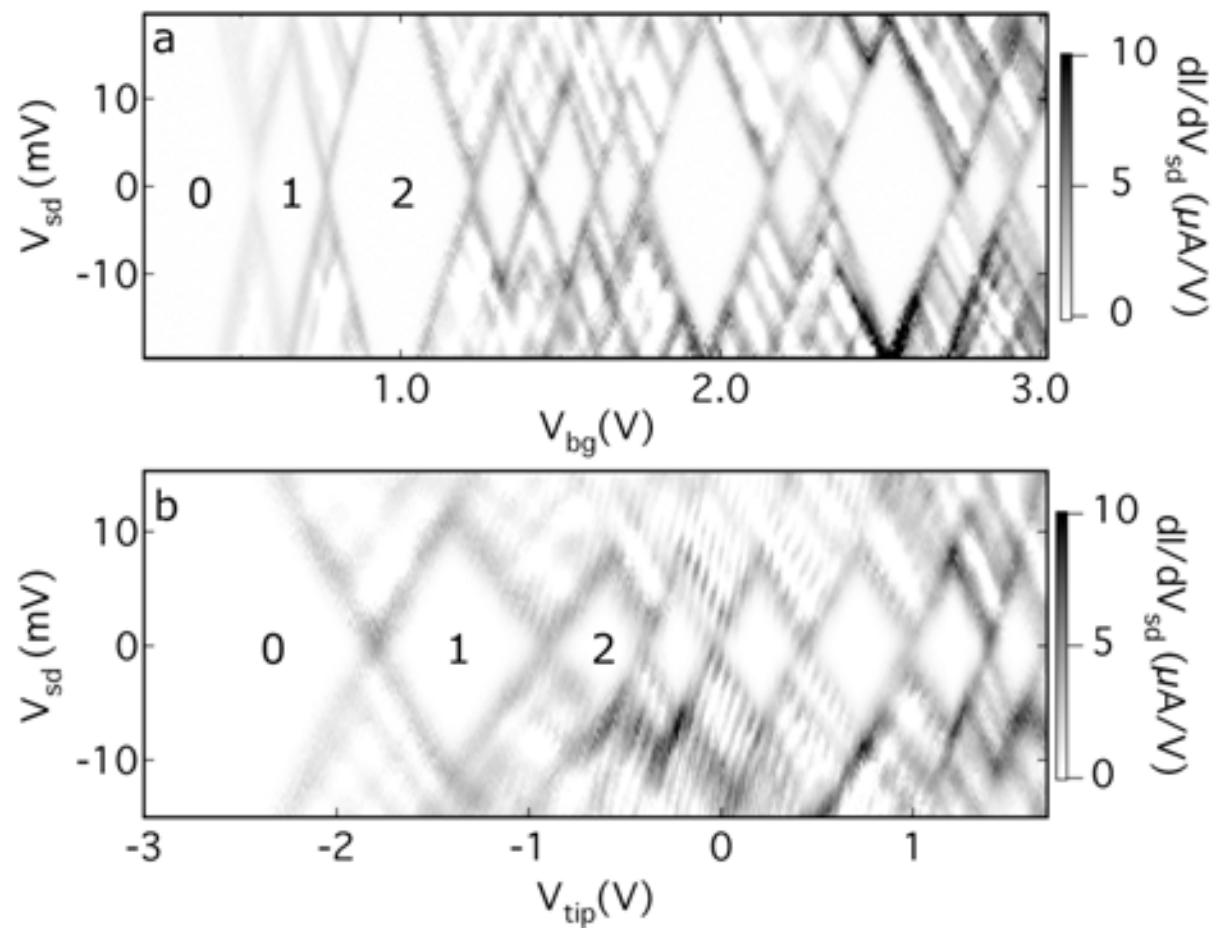

FIG. 2 Coulomb blockade diamonds for an InAs quantum dot in an InAs/InP nanowire heterostructure, plotting differential conductance *vs.* $V_{sd}$ and either (a) back gate voltage $V_{bg}$ or (b) tip voltage $V_{tip}$. These data show that either the back gate or the SPM tip can be used to tune the dot's charge. The differing size of the diamonds indicates the atomic-like shell structure of the quantum dot.  Both the tip and the back gate can reduce the number of electrons on the dot to 0 or 1 as indicated by the large area of zero conductance at the left of the two diamond plots.

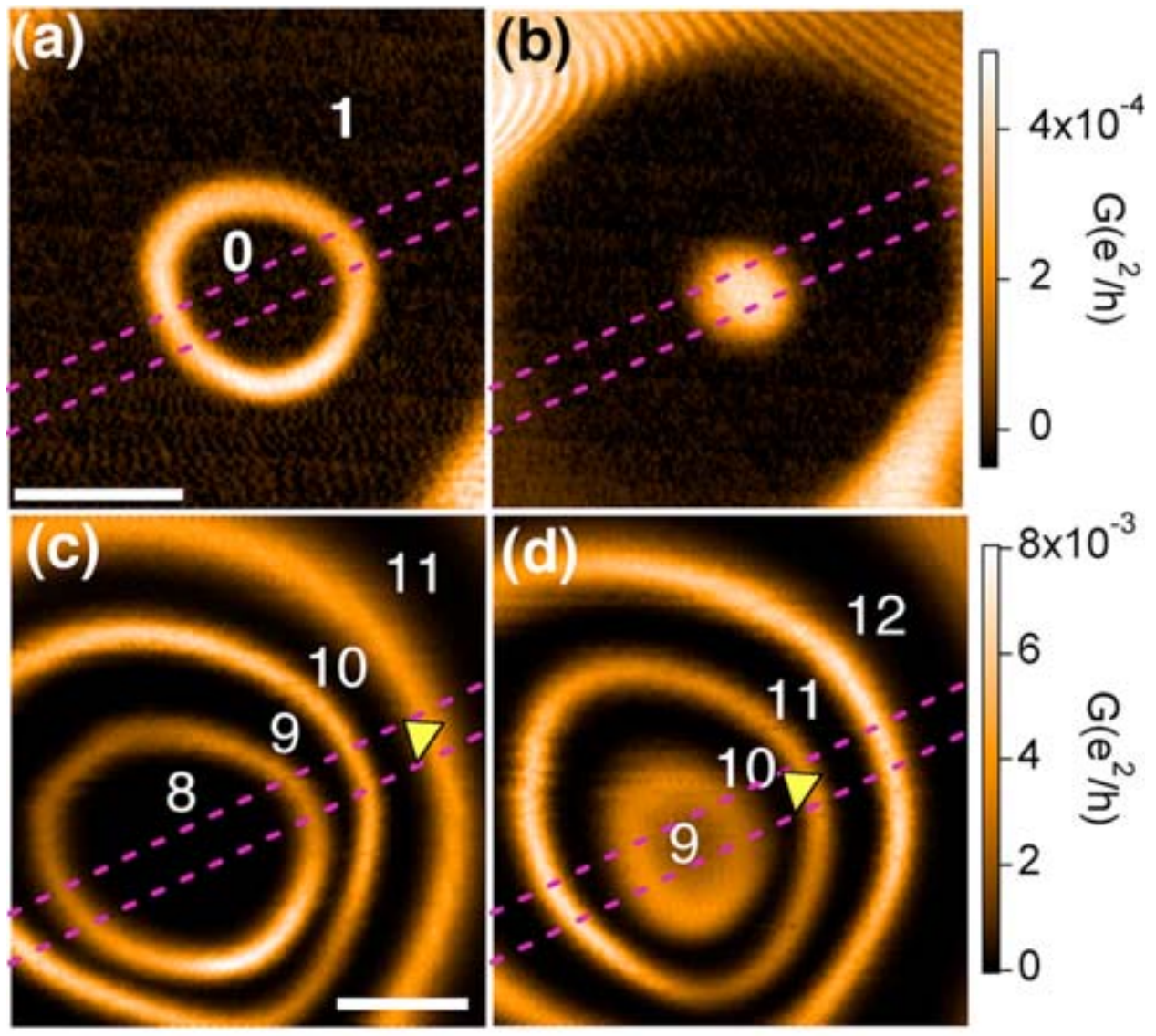

FIG. 3 SPM Coulomb blockade conductance images (a)-(b) of the last electron on the InAs nanowire quantum dot, and (c)-(d) of a higher number of electrons vs. tip position, as the tip is scanned in a plane 100 nm above the dot. The rings of high conductance correspond to Coulomb blockade conductance peaks. The integers indicate the number of electrons on the dot when the tip lies inside or between the rings. The tip and backgate voltages are (a) $V_{tip}$ = -2.5 V, (b) $V_{tip}$ = -1.5 V. (c-d) $V_{tip}$ = -2.0 V. The back gate voltage for (a-b) is 0.43 V. As the back gate voltage is increased from (c) $V_{bg}$ = 2.5 V to (d) 2.7 V, more electrons are added to the dot; the yellow triangle tracks the conductance peak corresponding to addition of the eleventh electron. The scale bar lengths are (a) 100 nm and (c) 200 nm.

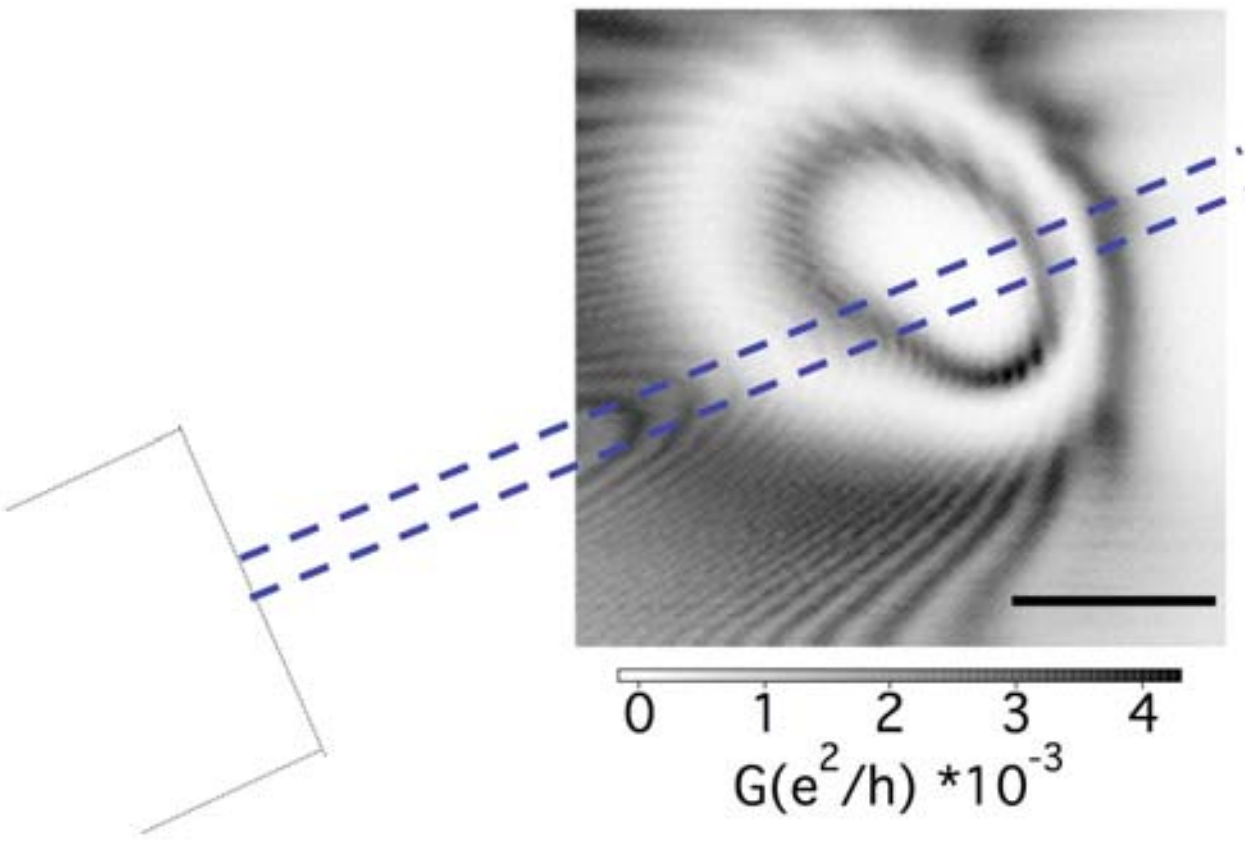

FIG. 4 SPM Coulomb conductance image obtained with tip voltage $V_t$ = 0.25 V and gate voltage $V_{bg}$ = 1.8 V. In addition to the rings surrounding the intentionally defined quantum dot, a second set of rings is seen in the lower left of the image. The extra set is centered another section of the nanowire, indicating that an extra dot has formed. The closer spacing of the second set of rings as well as their elongated shape indicate that the extra dot is much longer than the 18 nm defined dot. The scale bar is 200 nm.

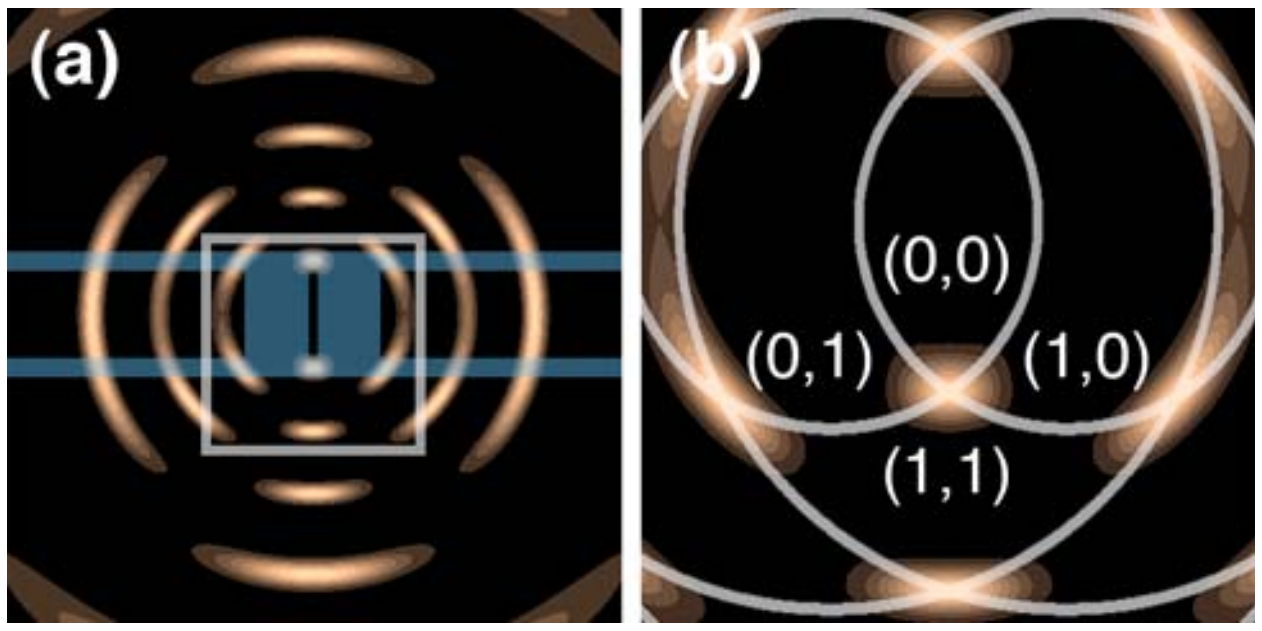

FIG. 5 SPM conductance image simulations of a double quantum dot formed in a 50nm diameter InAs/InP nanowire (outlined in blue). Dark (light) regions correspond to low (high) conductance. (a) The two InAs quantum dots (shaded in blue) are each 25 nm long and are defined by 5 nm thick InP tunnel barriers. The tip is scanned in a plane 25 nm above the nanowire. (b) Zoom-in of the boxed area in (a). The number of electrons on the left and right dot when the tip lies in that region are indicated.